\documentclass[preprint2]{aastex}

\begin{document}
\title{The coronal topology of the rapidly rotating K0 dwarf, AB Doradus 
I. Using  surface magnetic field maps to model the structure of the 
stellar corona }
\author{ G.A.J. Hussain, A.A. van Ballegooijen }
\affil{ Harvard Smithsonian CfA MS-16, 60 Garden Street, 
Cambridge MA 02138, USA}
\authoremail{ ghussain@cfa.harvard.edu, vanballe@cfa.harvard.edu }
\author{ M. Jardine, A. Collier Cameron }
\affil{ School of Physics and Astronomy, University of St. Andrews, 
North Haugh, St Andrews KY16 9SS, UK }
\authoremail{ mmj@st-andrews.ac.uk, acc4@st-andrews.ac.uk }

\begin{abstract}
We re-analyse spectropolarimetric data of AB Dor taken in 1996 December
using a surface imaging code that can model the magnetic field of the
star as a non-potential current-carrying magnetic field. 
We find that a non-potential field needs to be introduced in order
to fit the dataset at this epoch. This non-potential component
takes the form of a strong unidirectional azimuthal field 
of a similar strength to the radial field. This azimuthal field  
is concentrated around the boundary of the dark polar spot recovered at the 
surface of the star using Doppler imaging. As polarization signatures from 
the center of starspots are suppressed, it is unclear whether or not 
this non-potential component genuinely represents electric current
at the unspotted surface or whether it results from the preferred detection
of horizontal field in starspot penumbrae.
This model contains 20\% more energy than the corresponding potential field
model at the surface. 
This amount of free energy drops to under 1\% about 1{\,\mbox{$\mbox{R}_*$}}\
 above the photosphere.

We use these surface maps to model the coronal structure of the star.
The mixed radial polarities at the pole in the surface maps support
closed coronal loops in the high latitude regions, indicating that 
a component of the X-ray emission may originate in this area.
Assuming that the field remains closed out to 5{\,\mbox{$\mbox{R}_*$}}\,
we find stable surfaces where prominences may form out to the 
observed distances using this coronal model. 
\end{abstract}

\section{Introduction}

Doppler imaging 
allows us to map surface brightness distributions across the surfaces
of rapidly rotating stars (Vogt \& Penrod 1983).
This technique has proven to be an important tool when 
studying activity in rapidly rotating solar-type stars.
Spot maps show that, while {\em sunspots} 
tend to congregate between $\pm 30${\mbox{$^\circ$}}\ latitude,
spot patterns on other cool stars are different.
In rapid rotators, strong spot activity is often  concentrated 
in polar or high-latitude regions (Strassmeier 1996).
This difference to the solar pattern has been the subject of much debate
in recent years. The presence of strong flux at high latitudes  may
be explained in terms of  increased Coriolis forces acting on flux tubes
in the convective interior in rapidly rotating stars.
These would lead to the emergence of flux tubes in high latitude regions 
({Sch\"ussler} \& Solanki 1992).
An alternative  explanation for the polar spot phenomenon is 
that there are rings of alternating polarity at the pole 
(Schrijver \& Title 2001). 
These are produced when increased flux is injected into solar-type models. 
Meridional flows transport this flux to the pole.
As the flux injection rate is increased,
flux cancellation occurs more slowly and results in
strong flux gathering at the stellar pole.

In general, these dark starspots  are assumed to mark
the regions where the strongest magnetic flux is concentrated. 
However, 
it is likely that smaller spots exist below our resolution limit 
that we cannot reconstruct. 
 Semel (1989) proposed applying standard  Doppler imaging principles
to circularly polarized spectra in order to detect  magnetic fields on the 
surfaces of rapidly rotating stars. 
This technique is called Zeeman Doppler imaging  (ZDI).
Brown et al. (1991) first developed a code that employed maximum entropy
techniques to enable the mapping of magnetic fields on 
the stellar photospheres. 
This method has since been applied to three rapidly rotating systems:
the RS CVn binary, HR1099 (K1IV+G5V);  the K0 dwarfs; AB Dor and LQ Hya 
(Donati \& Collier Cameron 1997, Donati 1999).
The technique is not very sensitive to magnetic fields in 
spotted regions, but the resulting maps typically show the presence of 
strong magnetic field even in relatively unspotted parts of the photosphere.
A puzzling feature in these maps is that the radial and azimuthal field 
components across the surface are of about the same strength.
This is very different to the solar case, where strong horizontal field
is only found in sunspot penumbrae. 

A criticism of ZDI has been 
that no relationship is assumed between the three components of the field vector
thus enabling the reconstruction of physically unrealistic magnetic 
field distribution patterns such as  monopoles at the stellar surface. 
Potential field configurations and their extrapolations have  
been used to model the global surface and 
coronal field of the Sun  (Altschuler \& Newkirk 1969). 
We show how the surface and coronal field of AB Dor can be modeled 
using the magnetic field maps obtained using an advanced version of ZDI. 
Hussain, Jardine \& Collier Cameron (2001) describe a technique that 
introduces a relationship between the different components of the 
surface magnetic field vector by assuming that the
observed data can be fitted using a potential field distribution. 
In this paper we extend this method to non-potential fields,
i.e. we assume
that the surface field can be 
modeled using a potential field plus a non-potential field component
that represents the presence of electric currents.
This method is similar to that described by  Donati (2001). 

The K0 dwarf, AB Doradus (HD 36705), is a relatively bright 
($m_{V}\approx 7$) example of a class of very active cool stars 
that are just evolving onto the main sequence. 
Its distance has been measured to be 15pc using HIPPARCOS and VLBI 
interferometry and its age is estimated to be between 20 to 30~Myr 
(Guirado et al. 1997, Collier Cameron \& Foing 1997).
It rotates at roughly 50 times the solar rate, 
$P_{\mbox {rot}}=0.51479$~d (Pakull 1981). 
Its surface is covered with large cool starspots as indicated by 
rotational modulation of its photometric light curve (Rucinski 1985).
AB Dor also displays strong X-ray variability on all timescales 
({K\"urster} et al. 1997). 
Long-term photometry indicates that  the
star was at its brightest in the first epoch of observation in 1978,
it then decreased down to a minimum level in 1988 and has since 
been increasing steadily, most recently plateauing out to roughly
the same $m_V$ level as 1978.
This may be evidence of a solar-type activity cycle with
the starspot coverage increasing and decreasing with a period of 22-23 years
(Amado 2001).

Doppler images of this star have revealed the presence of a 
polar spot that has been consistently present since 1992 
(Collier Cameron \& Unruh 1994, Collier Cameron 1995, 
Donati \& Collier Cameron 1997, Donati et al. 1999).
The only image of AB Dor that does not display this polar feature was 
obtained using data taken in 1989 ({K\"urster}, Schmitt \& Cutispoto 1994),
the star was at its minimum brightness level at this epoch and
hence at its most spotted (Amado 2001).
Hence this feature may be associated in some way with a stellar activity 
cycle. This polar spot should be associated with strong magnetic field
if the solar stellar analogy applies.
Lower-latitude spots are also recovered in Doppler maps but are found
to be much less stable.

ZDI studies of AB Dor carried out annually since 1995
reveal magnetic field maps with strong field in even
relatively unspotted parts of the photosphere.
The strong unidirectional azimuthal  flux recovered near the pole
in maps obtained in successive years provides a puzzling non-solar
pattern (Donati \& Collier Cameron 1997, Donati et al. 1999, Hussain 2000). 
We re-analyse the best sampled dataset using a code that 
introduces a relationship between all three field components,
and use this analysis to evaluate if this band of azimuthal flux is
still present. These new surface maps are extrapolated
to model the structure of the corona. This coronal topology
is used to evaluate general properties of the stellar corona.
The model is also used to compute sheet-like surfaces of
stable mechanical equilibria taking the rotation of the star into account
(Jardine et al. 2001). 
These are sites where the prominences are likely to form.
We use these coronal magnetic field models to model the temperatures 
and densities expected in the corona of AB Dor in paper II 
(van Ballegooijen \& Hussain 2002).

\section{Magnetic field mapping}
\label{sec:technique}
The basic principles behind the technique of Zeeman Doppler imaging (ZDI)
are presented here (Semel 1989; Donati \& Coller Cameron 1997). 
Circularly polarized spectra, ``Stokes V spectra'', are sensitive
to the {\em line-of-sight components} of the magnetic field. 
In the case of high resolution spectra from rapidly rotating stars 
the circularly polarized contributions from different regions of the 
visible stellar disk are separated in velocity-space. 
As the star rotates, a time series of circularly polarized
spectra allow us to pinpoint the {\em locations} of magnetic field regions. 
High latitude magnetic field areas produce distortions that are
confined to the centre of the line profile while low-latitude magnetic
field distortions move out into the wings of the profile.

Stokes V spectra are sensitive to the longitudinal component of the 
field. Within the weak field regime (applicable to magnetic field strengths 
under 1kG) the amplitude of the signature scales with  surface area 
covered by the magnetic field, and with field strength.
Different field orientations produce distinctive signatures in dynamic
spectra. Hence a time-series of spectra allows us to ascertain 
both the {\em size}, {\em strength} and 
the {\em orientation} of the magnetic region (Donati \& Brown 1997).

This technique has been used to derive surface magnetic field maps for 
the K0 dwarves AB Dor, LQ Hya and the active G dwarf in the
RS CVn binary system HR1099 
(Donati \& Coller Cameron 1997; Donati et al. 1999; Donati 1999).
The magnetic field has been reconstructed assuming 
no interdependence between 
the radial, azimuthal and meridional field
maps. This can lead to monopoles and other physically 
unrealistic magnetic field distributions.
In the case of high inclination stars like AB Dor ($i=60^{\circ}$), 
the circularly polarized 
signatures from radial and
meridional magnetic field regions are very similar and thus lead to 
cross-talk (Donati \& Brown 1997).
By imposing a relationship between the different components of the 
magnetic field vector, we can compensate for
limitations in ZDI caused by the lack of sensitivity of 
circularly polarized profiles to certain field orientations such as 
low-latitude meridional field vectors 
(Hussain, Jardine \& Collier Cameron 2001).

We have developed an advanced version of ZDI that imposes a
relationship between the three components of the vector field
${\bf B} ({\bf r})$. The magnetic field is written as a linear
combination of non-potential and potential fields:
\begin{equation}
{\bf B} ({\bf r}) = {\bf B}_{\rm np} ({\bf r}) - \nabla D ,
\end{equation}
where $\nabla \cdot {\bf B}_{\rm np} = 0$ and $\nabla^2 D = 0$.
A simple model for the non-potential field is used. First, we assume
that the radial component of the non-potential field vanishes,
$B_{{\rm np},r} = 0$. This is true by definition at the stellar
surface, but we assume here for simplicity that it is also true
at larger heights. Since $\nabla \cdot {\bf B}_{\rm np} = 0$,
the tangential components can be written as:
\begin{equation}
B_{{\rm np},\theta} = \frac{1}{r \sin \theta} \frac{\partial F}
{\partial \phi} , ~~~~~~
B_{{\rm np},\phi} = - \frac{1}{r} \frac{\partial F} {\partial
\theta} , \label{eq:Bnp}
\end{equation}
where $F({\bf r})$ is a scalar. This implies that the field lines
of the non-potential field are closed curves that lie on spherical
shells ($r$ = constant, $F$ = constant). Second, we assume that
the electric current density ${\bf j} ({\bf r})$ is derived from
a potential $Q({\bf r})$:
\begin{equation}
\frac{4 \pi}{c} {\bf j} = \nabla \times {\bf B}_{\rm np} =
- \nabla Q . \label{eq:curr}
\end{equation}
It follows that $\nabla^2 Q = 0$, so the distribution of electric
currents in the corona is uniquely determined by the current density
$j_r (R_*, \theta, \phi)$ through the photosphere. Inserting equation
(\ref{eq:Bnp}) into equation (\ref{eq:curr}), we find that $F$ can be
written as $F \equiv r^2 \partial C / \partial r$, where $C({\bf r})$
is a harmonic function ($\nabla^2 C = 0$) and $Q$ is given by
\begin{equation}
Q \equiv \frac{1}{\sin \theta} \frac{\partial} {\partial \theta}
\left( \sin \theta \frac{\partial C} {\partial \theta} \right)
+ \frac{1}{\sin^2 \theta} \frac{\partial^2 C}{\partial \phi^2}
= - \frac{\partial}{\partial r} \left( r^2 \frac{\partial C}
{\partial r} \right) .
\label{eq:Q}
\end{equation}
The above assumptions are rather arbitrary, but they are an
improvement over the conventional ZDI method in that they allow us to
reconstruct electric currents {\it in addition to} the potential field.

The magnetic field is extrapolated out to a source surface, $R_s$,
beyond which the field is assumed to be radial 
(Schatten, Wilcox \& Ness 1969). 
While this solution is not force-free it allows us to
pinpoint the areas where the surface field distribution becomes
non-potential while still allowing for the extrapolation of the field out
to the corona.
The magnetic field is expressed in terms of spherical harmonics:

\begin{figure*}
\epsscale{1}
\plotone{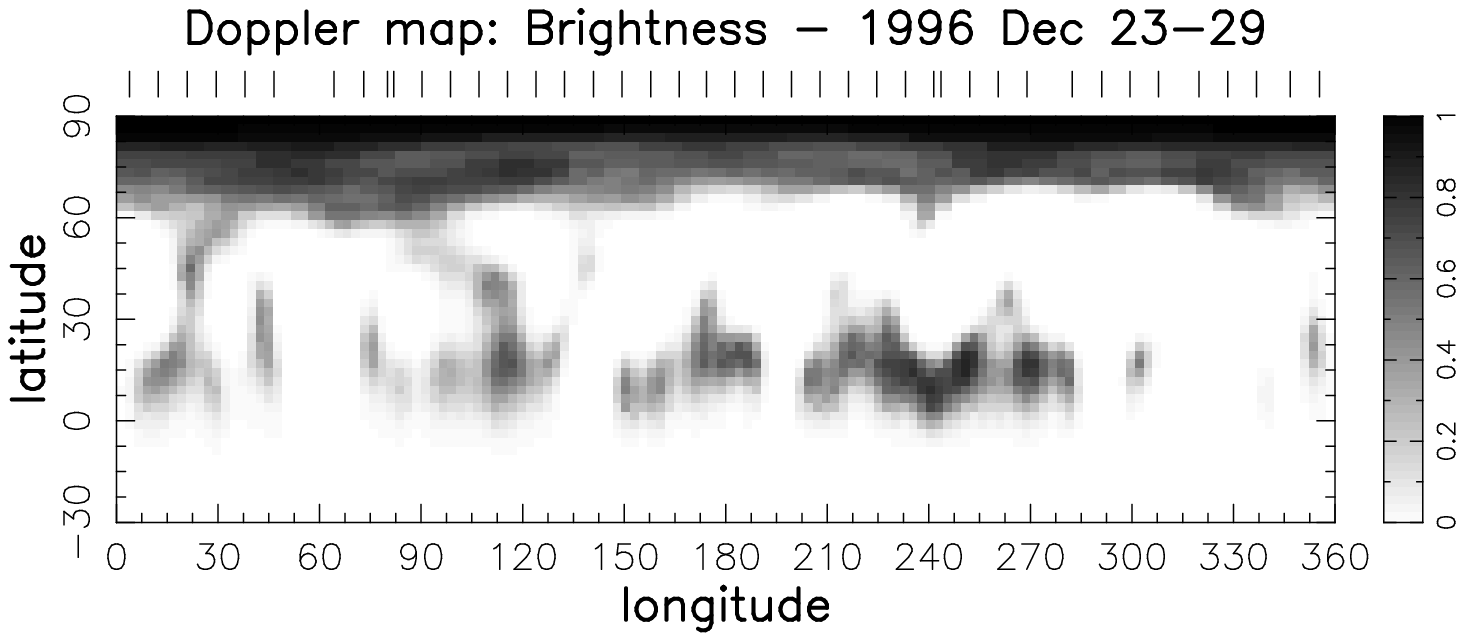}
\caption{Surface brightness map for AB Dor in 1996 Dec 23-29.
The greyscale for the brightness 
map represents spot occupancy for the star (1=complete spot coverage).
The tick marks above each image indicate each phase of observation.}
\label{fig:brimap}
\vspace{5mm}
\epsscale{2.0}
\plotone{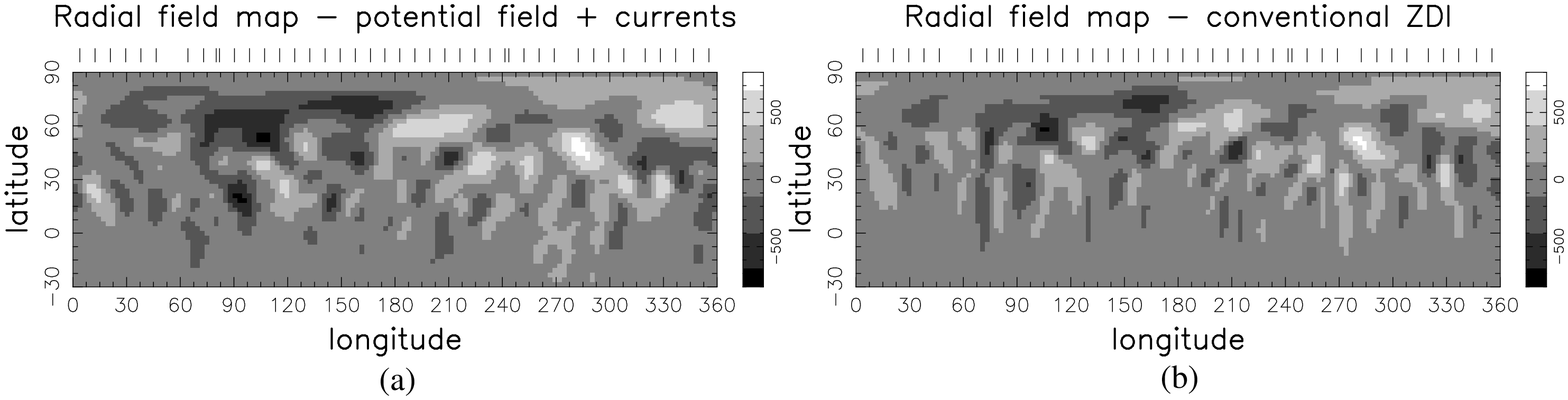}

\vspace{2mm}

\plotone{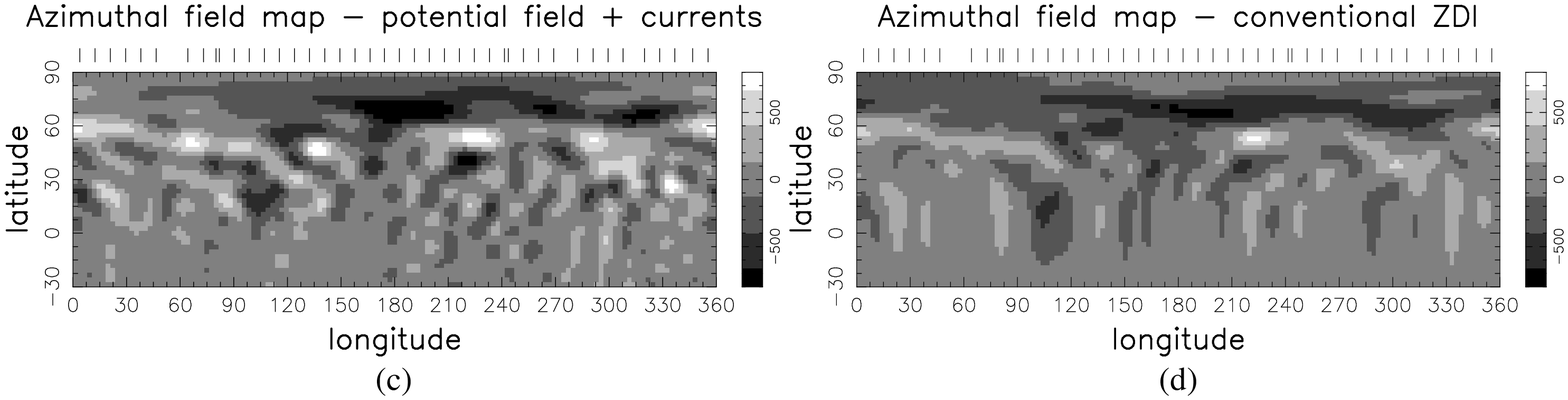}

\vspace{2mm}

\plotone{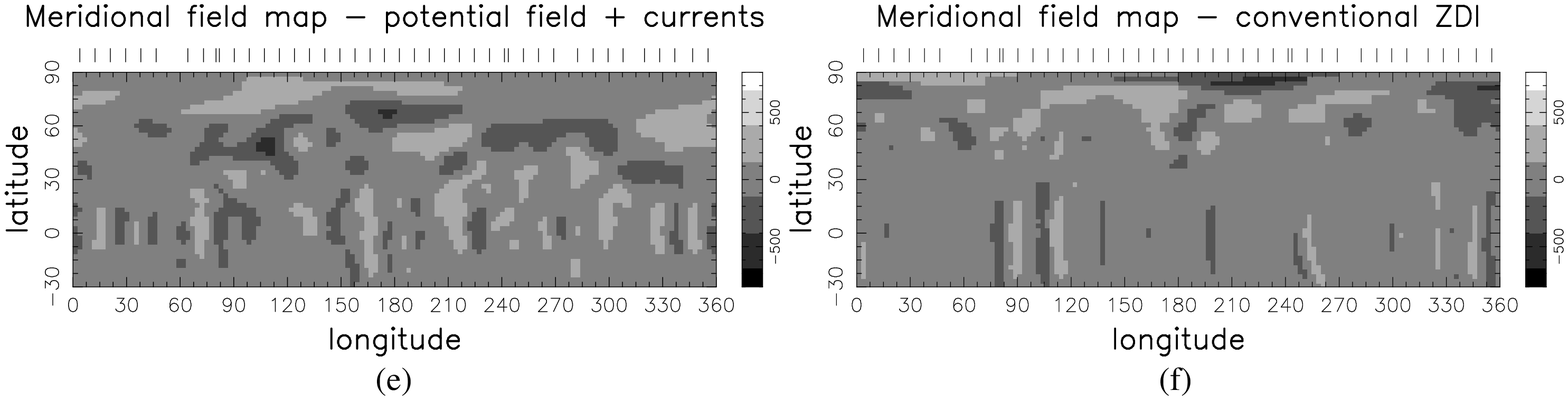}
\caption{Magnetic field maps for AB Dor in 1996 Dec 23-29.
The magnetic field maps in the left column were produced
using the potential field and current prescription described here. 
The maps in the right column are those obtained using conventional ZDI 
(i.e. no assumed relationship between the three components
of the magnetic field vector).
All greyscales are the same (white/black represent $\pm$~800G respectively). 
Both models fit the observed spectra to the same degree (reduced $\chi^2$=1.1).
}\label{fig:magmaps}
\end{figure*}

\begin{figure*}
\epsscale{2.0}
\plotone{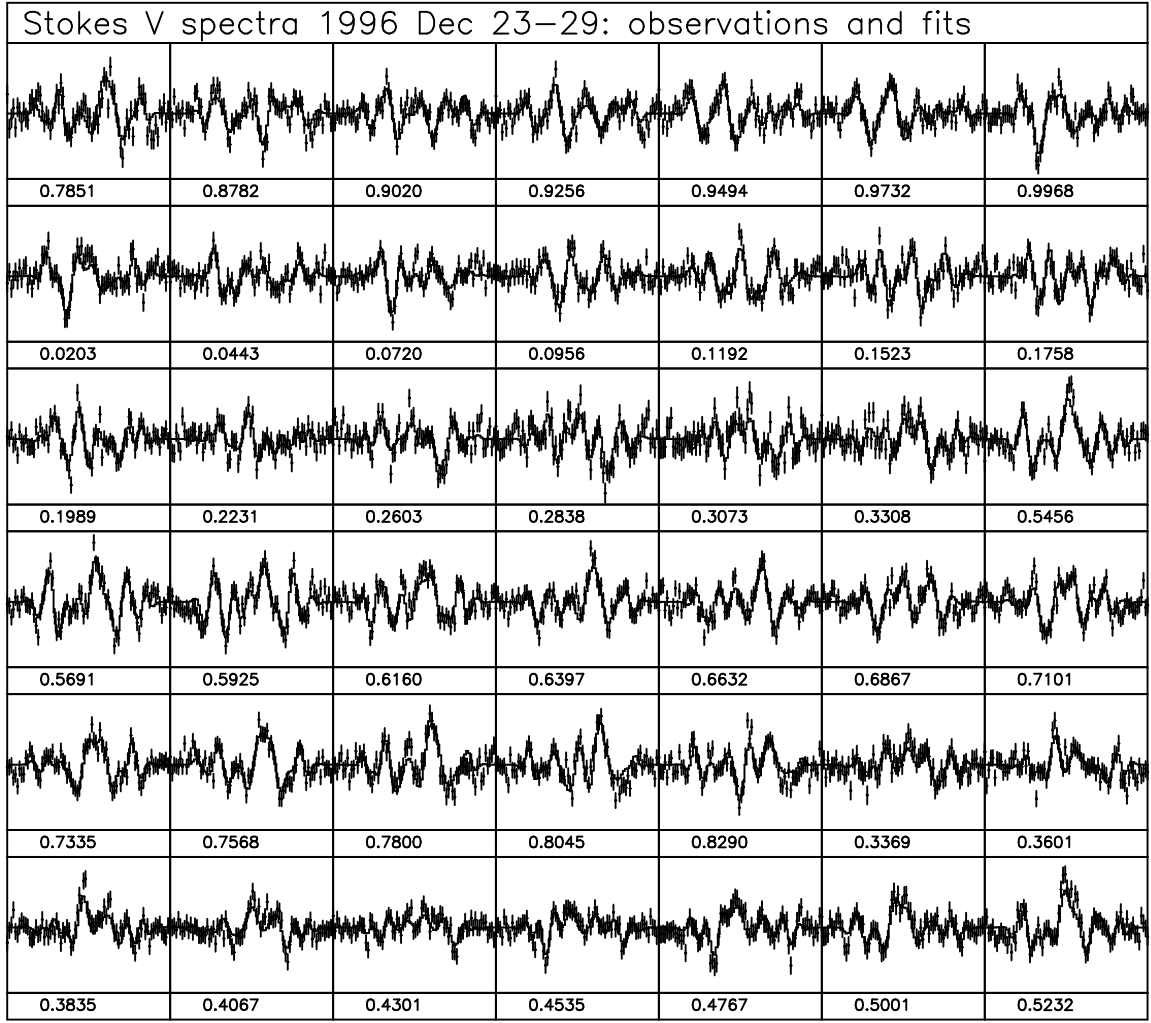}
\caption{The observed Stokes V intensity spectra from 1996 Dec 23-29 
are shown by the error bars.
The solid lines represent the spectral line profiles based on the non-potential
magnetic field model (shown in the left-hand column of Fig.2).
All spectra are plotted on the same intensity scale;
the maximum polarization is $\pm 0.08$\% of the continuum level.
The phase of each observation is denoted at the bottom of each spectrum.}
\label{fig:specfits}
\end{figure*}

\begin{eqnarray}
B_{r}(r,\theta,\phi) & = & \sum^{N}_{l=1} \sum^{l}_{m=-l} B_{lm}P_{lm}(\theta)f_{l}(r)e^{im\phi} , \\
B_{\theta}(r,\theta,\phi) & = & - \sum^{N}_{l=1} \sum ^{l}_{m=-l} [ B_{lm}\frac{d}{d\theta}P_{lm}(\theta)g_{l}(r) \nonumber \\
& & + (l+1) C_{lm} \frac{P_{lm}(\theta)}{\sin\theta}im r^{-(l+1)}] e^{im\phi}, \\
B_{\phi}(r,\theta,\phi) & = & - \sum^{N}_{l=1} \sum ^{l}_{m=-l} [ B_{lm}\frac{P_{lm}(\theta)}{\sin\theta} im g_{l}(r) \nonumber \\
& & - (l+1) C_{lm} \frac{d}{d\theta}P_{lm}(\theta)r^{-(l+1)}]e^{im\phi} .
\label{eq:radazmer}
\end{eqnarray}

\noindent
Here $\theta$ represents co-latitude; $\phi$ represents longitude;
$B_{lm}$ represents the radial field at the stellar surface;
$C_{lm}$ is the coefficient of the electric current;
$P_{lm}(\theta)$ is the associated Legendre function at each latitude; 
the functions, $f_l(r)$ and $g_l(r)$ are given by 
(Nash, Sheeley \& Wang 1988):
\[ f_{l}(r)=\frac{(l+1)r^{-l-2}+lR_{s}^{-2l-1}r^{l-1}}{l+1+lR_{s}^{-2l-1}}, \] 
\[ g_{l}(r)=\frac{r^{-l-2}-R_{s}^{-2l-1}r^{l-1}}{l+1+lR_{s}^{-2l-1}}. \]

We use \( N=23 \) as this corresponds to a spatial resolution of 
2.5$^{\circ}$ at the equator (i.e. the limit of the spectral resolution).
The ZDI code maps different $B_{lm}$ and $C_{lm}$ configurations until the
dataset is fitted. The radial, azimuthal and meridional field
maps at the surface are then reconstructed using Eqns.~(5), (6) and (7).
The source surface, $R_s$ is initially set to 
5{\,\mbox{$\mbox{R}_*$}}\ as prominence-type complexes are found to be 
supported out to these heights at this epoch (Donati et al. 1999).
See Secn.~\ref{sec:extrapol} for a more detailed 
discussion of this parameter.

\subsection{Weighting scheme}

As several solutions can fit the observed spectral dataset to within
any level of $\chi^{2}$, a regularising function is required to obtain 
a unique solution. We use maximum entropy in order to obtain the 'simplest'
possible image. 
We define the simplest image as the image that has
the least number of nodes across
the stellar surface while still fitting the observed dataset.
Weights are assigned such that the image is biased towards 
the lowest order harmonics. 
See Hussain, Jardine \& Collier Cameron (2001) 
for more discussion on the effect of different weighting schemes.
The weighting scheme  used  here  defines weights as follows:

\begin{equation}
\frac{1}{W(l,m)}= [(2|m| + 1)(l-|m|+1)]^{2}.
\label{eq:weights}
\end{equation}

\noindent
Different schemes were tested and this one was chosen as it 
produced the quickest convergence to a unique solution.

\subsection{Dataset}

The spectroscopic observations used to obtain the magnetic field
map for AB Dor were obtained 
as part of a long-term stellar activity monitoring 
programme over the week of 1996 December 23-29.
The Semel polarimeter (Semel 1993)
was mounted at the Cassegrain focus of the 3.9m Anglo-Australian 
Telescope and linked to the {\mbox{\sc ucles}}\ spectrograph.
For details of the instrumental setup and observing run, see
Donati et al. (1997) and Donati \& Collier Cameron (1997).
The technique of {\em least squares deconvolution} (LSD) is used to 
improve the signal-to-noise ratio of the Stokes V profiles
(Donati et al. 1999).
It essentially cross-correlates thousands
of selected line profiles with an appropriately weighted
mask to produce a mean line profile with reduced noise. 

The intrinsic line profile is assumed to remain 
constant across the entire visible stellar disc. 
The central wavelength and Land\'e factor of the mean LSD line profile 
is used to generate a look up table that contains the
mean Stokes V  profiles as a function of limb angle
for a magnetic field at the equipartition
strength (taken to be a maximum at 1.5~kG). 
See Hussain et al. (2000) for more detail on how these LSD
profiles were produced.
Data were combined from all four nights (1996 Dec. 23, 25, 27 and 29) in
order to obtain as complete phase coverage as possible.

\section{Surface magnetic field map of AB Dor}

Initially the observed circularly polarized spectra were analysed using 
a code that just allowed for potential field solutions 
(Hussain, Jardine \& Collier Cameron 2001). 
The observed differential rotation rate of the star, $\Omega(l)$, 
was taken into account when combining spectra taken over 1 week 
(i.e. 14 rotation cycles):
$\Omega(l) = 12.2434-0.0564 \sin^2 l$ rad~ d$^{-1}$, where $l$ is the latitude
on the star (Donati \& Collier Cameron 1997, Donati et al. 1999).
However, it was not possible to
find a solution with a reduced $\chi^2 < 2 $ for this dataset.

On allowing for an electric current component  we find that this
provides the additional freedom to fit the 
observed dataset to a reduced $\chi^2$ level of 1 (see Fig.~\ref{fig:specfits}). 
It should be noted that this is not a force-free solution.
However, this method does have the advantage in that it does allow us 
to pinpoint  the regions at which the field distribution departs 
from being purely potential, 
as well as enabling us to extrapolate the surface field out to the corona.

As AB Dor is inclined at $i=60${\mbox{$^\circ$}}\ the 
magnetic field cannot be reconstructed reliably below a latitude,
$l=-30${\mbox{$^\circ$}}. 
The high inclination angle means that the contribution of the 
meridional (N-S) field component to the 
circularly polarized profiles is  suppressed and subject to 
cross-talk with the radial field component, especially at lower latitudes
 (Donati \& Brown 1997). 
This may explain why the recovered meridional field is 
considerably weaker than either radial or azimuthal field components.
There may be stronger meridional
field at the stellar surface that we cannot recover using  
circularly polarized profiles alone.
Once Stokes Q and U profiles can also be included in the reconstruction
process it should be possible to improve the accuracy of the meridional field
maps.
Unfortunately at present it is not possible
to obtain linear polarization signatures with sufficiently
high S/N for this purpose.

The magnetic field analysis assumes that the intrinsic line depth 
across the stellar surface is uniform. 
This means that the spotted regions remain unaccounted for and 
the magnetic field at the surface is likely to be much stronger,  
particularly in the dark, spotted regions.
This especially applies to the polar cap, where the dense
spot coverage  suppresses the contribution to the polarization
signature from the entire region. 
It is possible that the strong azimuthal field that is recovered
represents strong horizontal field in the penumbrae of the starspots
and that more radial field at the surface is suppressed as it
is located in the center of the spot umbrae. 
While the contribution from the meridional (N-S) field component
to circularly polarized spectra is suppressed in the lower latitudes 
of  high inclination stars (like AB Dor),
the meridional field contribution should be strong in the polar region.
The origin of the strong negative azimuthal field in the 
penumbrae of the polar region  is still unclear.

\subsection{Comparison with conventional ZDI}

The maps shown in Fig.~\ref{fig:magmaps} are Cartesian projections
of the stellar surface. 
The radial and azimuthal magnetic field maps in Fig.~\ref{fig:magmaps}
show very similar distributions to those obtained using conventional ZDI 
(i.e. no assumed relationship between the components of the field vector).
The radial field maps (Figs.~\ref{fig:magmaps}a~\&~b) differ
only in the relative areas of the field that are reconstructed.
These differences are not significant and can be affected by slight
differences in the degree of fit that the images are pushed to 
(i.e. the level of agreement to the data as measured by $\chi^{2}$).
Both the non-potential magnetic field model and the ZDI model fit the observed
spectra to a reduced $\chi^2=1.1$. Fig.~\ref{fig:specfits} shows the fits to
the observed spectra from the  non-potential magnetic field model.
There is some  difference between the two azimuthal field
maps (Figs.~\ref{fig:magmaps}c~\&~d).
In particular, in the conventional ZDI model there is an unbroken band
of negative azimuthal field above a latitude of 
60{\mbox{$^\circ$}}\, whereas in the non-potential field model
some positive azimuthal field is also reconstructed near longitude 
0{\mbox{$^\circ$}}.

As with Doppler imaging, spectral signatures from 
low latitude features move through the line profile much more quickly than
the signatures from higher latitude fields as the star rotates.
This makes it harder to constrain the location of equatorial features.
As the code preferentially produces images with as little structure as possible,
it essentially weights against equatorial features which have large areas.
Structure is pushed away from the equator and this results in the smearing
effect seen in these maps.  This smearing is particularly
noticeable in the meridional field maps, where low-latitude spots are
subject to cross-talk with the radial field maps in high inclination stars.
The lack of constraints in the meridional field maps is also reflected in 
the differences between Figs. ~\ref{fig:magmaps}e~\&~f.

\begin{figure}
\epsscale{1}
\plotone{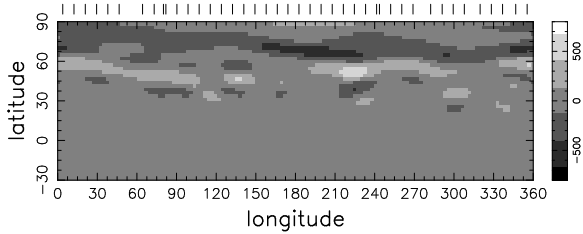}
\caption{Non-potential component of azimuthal field for the same reconstruction
as shown in Fig.2c.
Greyscales and tick marks represent the strength of the magnetic field 
(in Gauss) and observation phases respectively 
(as in Fig.~\ref{fig:magmaps}).}
\label{fig:curmaps}
\end{figure}

\subsection{Azimuthal component}  

As mentioned above, 
these new maps do not show the same unbroken band of azimuthal field near the pole. 
While there is still strong negative azimuthal field in this region, the
band is not complete and is even broken
with a region of positive polarity near 0{\mbox{$^\circ$}}\ longitude 
(see Fig.~\ref{fig:magmaps}c). 
Fig.~\ref{fig:curmaps} shows the contribution of the electric currents
(the last term in Eqn.~7) to the azimuthal field.
In this figure, there is some field around 50$^\circ$ latitude but
the  strongest field is associated with the strong negative band of 
azimuthal field between 70$^\circ$ and 80$^\circ$ latitude. 

Preliminary results obtained by carrying out a similar analysis
on 1995 and 1998 datasets also suggest that the strong negative 
component may be increasing in strength with subsequent years.
In the 1995 dataset, there is no need to invoke a non-potential
component to fit the observed spectra within a reduced $\chi^2\approx 1$.
What this non-potential component actually represents is unclear.
As these maps are flux-censored in the spotted regions, it may be that
the non-potential component exists only in certain parts of the penumbrae
and not in others. The fraction of the penumbrae with azimuthal fields is 
apparently increasing with time. The origin of this magnetic shear 
in the penumbra of the polar spot is still unclear.

\section{Coronal magnetic field }
\label{sec:extrapol}

\begin{figure*}
\epsscale{2.0}
\plotone{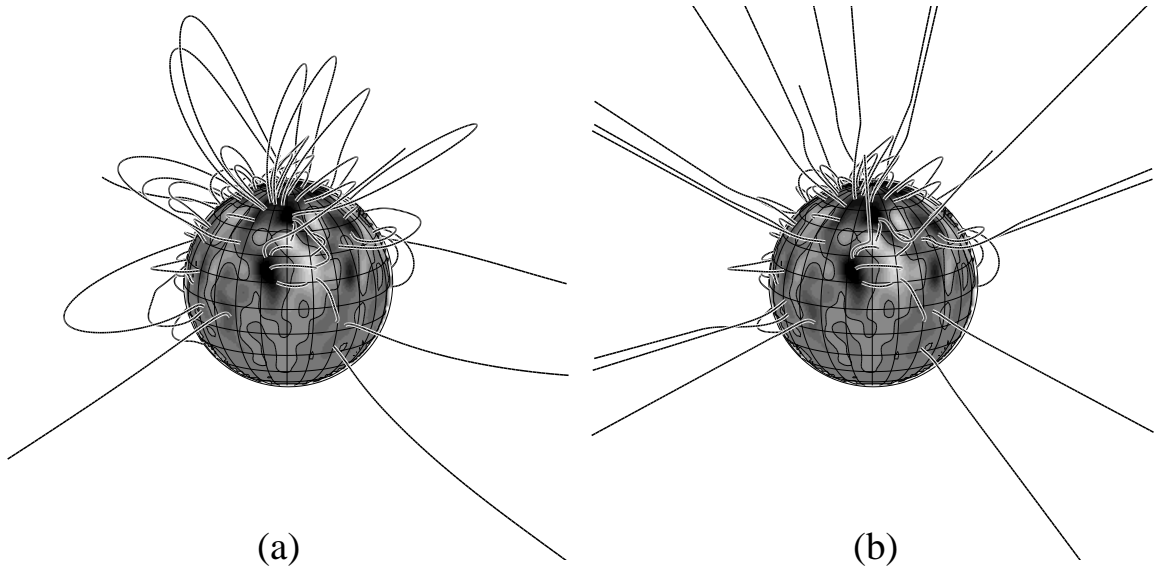}
\caption{Coronal magnetic field of AB Dor for two values of the source 
surface radius: (a) $R_s=5${\,\mbox{$\mbox{R}_*$}}, and 
(b) 1.7{\,\mbox{$\mbox{R}_*$}}.}
\label{fig:corfield}
\end{figure*}

As described in Secn.~\ref{sec:technique} the magnetic field, 
{\bf B}, can be written in terms of spherical harmonic
functions and can be used to determine the radial dependence 
of the field. The resulting coronal magnetic field 
topology is shown in Fig.~\ref{fig:corfield}. The magnetic field
is shown for two values of the source surface radius, $R_s$ 
[see Eqns.5-7].

Fast moving H$\alpha$ absorption transients have been reported
in the optical spectra of this object at every epoch at which it has been 
observed (Robinson \& Collier Cameron 1986, Collier Cameron et al. 1990).
These transients can be explained in terms of the presence of  
slingshot-type prominence complexes 
forced into rigid corotation with the system. The
rate of the transit of these cool (9000~K) clouds allow us to determine 
their axial distances from the stellar rotation axis.
In 1996 December these prominences were found at various distances
above the Keplerian corotation radius of the star 
($R_k \approx 2.6${\,\mbox{$\mbox{R}_*$}}) 
up to 5{\,\mbox{$\mbox{R}_*$}}\ (Donati et al. 1999).
This means that the magnetic field must be capable of supporting
closed loop structures at these distances and so we initially set 
$R_s = 5${\,\mbox{$\mbox{R}_*$}}\ (see Fig.\ref{fig:corfield}a).
However, in paper II we argue on the basis of the observed EUV emission
 from AB Dor that the gas pressure in coronal loops
with heights beyond 1.7{\,\mbox{$\mbox{R}_*$}}\
would exceed the magnetic presure,
hence the plasma in these loops cannot be magnetically contained.
Therefore in Fig.\ref{fig:corfield}b we show a second model with the source surface located 
at $R_s=$1.7{\,\mbox{$\mbox{R}_*$}}.

The mixed polarities in the surface radial field map allow 
closed loops to form over the polar region of the star.
This indicates that plasma can be supported over the pole of the star, 
and that at least some of the observed coronal emission from AB Dor
could originate here. 
This is indeed what is suggested by observations.
For example, Maggio et al. (2000) report an observation of a 
compact flare ($H \approx 0.3${\,\mbox{$\mbox{R}_*$}})
which took a rotation period
to decay and showed no evidence of self-eclipse. 
Furthermore, there is evidence from other active cool star systems, 
such as in the binaries 44i~Boo and Algol. In the former case, a strong
emission feature at $T\approx 8\times 10^6$~K indicating very dense, 
hot plasma confined close to the stellar surface. 
That this emission showed no sign of eclipse over a rotation cycle 
suggests that these hot loops may be located at the pole of the star
(Brickhouse \& Dupree 1998).
A flare observed in Algol (B8 V + K2 IV) using BEPPOSAX 
originating on the K component was completely eclipsed by the hot star
and allowed accurate determination of its physical characteristics.
The analysis revealed that the flare's maximum height was
0.6{\,\mbox{$\mbox{R}_*$}}\
and that it was located at the southern pole of the cool star component
of the system, Algol B (Schmitt \& Favata 1999).

\section{H$\alpha$ prominences}
\label{sec:proms}

As described in Secn.\ref{sec:extrapol}, there is evidence
for cool slingshot-type prominences formations lying well above
the Keplerian co-rotation radius of AB Dor, at distances of up to 5{\,\mbox{$\mbox{R}_*$}} (Donati et al. 1999).
Ferreira (2000) decribes how a prominence may be modeled in rapidly rotating
stars if it is represented as an axisymmetric 
current sheet in a perfectly conducting plasma. 
An equilibrium can be found between the gravity, centrifugal
and Lorentz forces acting on the current sheet. 
These prominences are most likely to form at stable points in the corona. 
Stable equilibria satisfy two conditions (Ferreira 2000):
\begin{itemize}
\item for equilibrium, the effective gravity component, $g_{\hbox{eff}}$, 
along a field line B must be  zero and
\item for a stable equilibrium, the location must be a minimum of the effective gravitational potential.
\end{itemize}
These stable points are also positions where the plasma pressure is a
maximum along a field line.  Jardine et al (2001) developed a general method
to compute the locations of these stable points in arbitrary field
configurations.  We have improved their technique by solving numerically
for positions of pressure maxima along field lines rather than
interpolating from a grid of field vectors.  This is more accurate close to
the star where the field is complex and avoids the problem with the
original method of a coarse grid leading to spurious stable points.  Here
we apply this method to the  non-potential
field model of AB Dor shown in Fig.~\ref{fig:corfield}a.

\begin{figure*}
\epsscale{2.0}
\plottwo{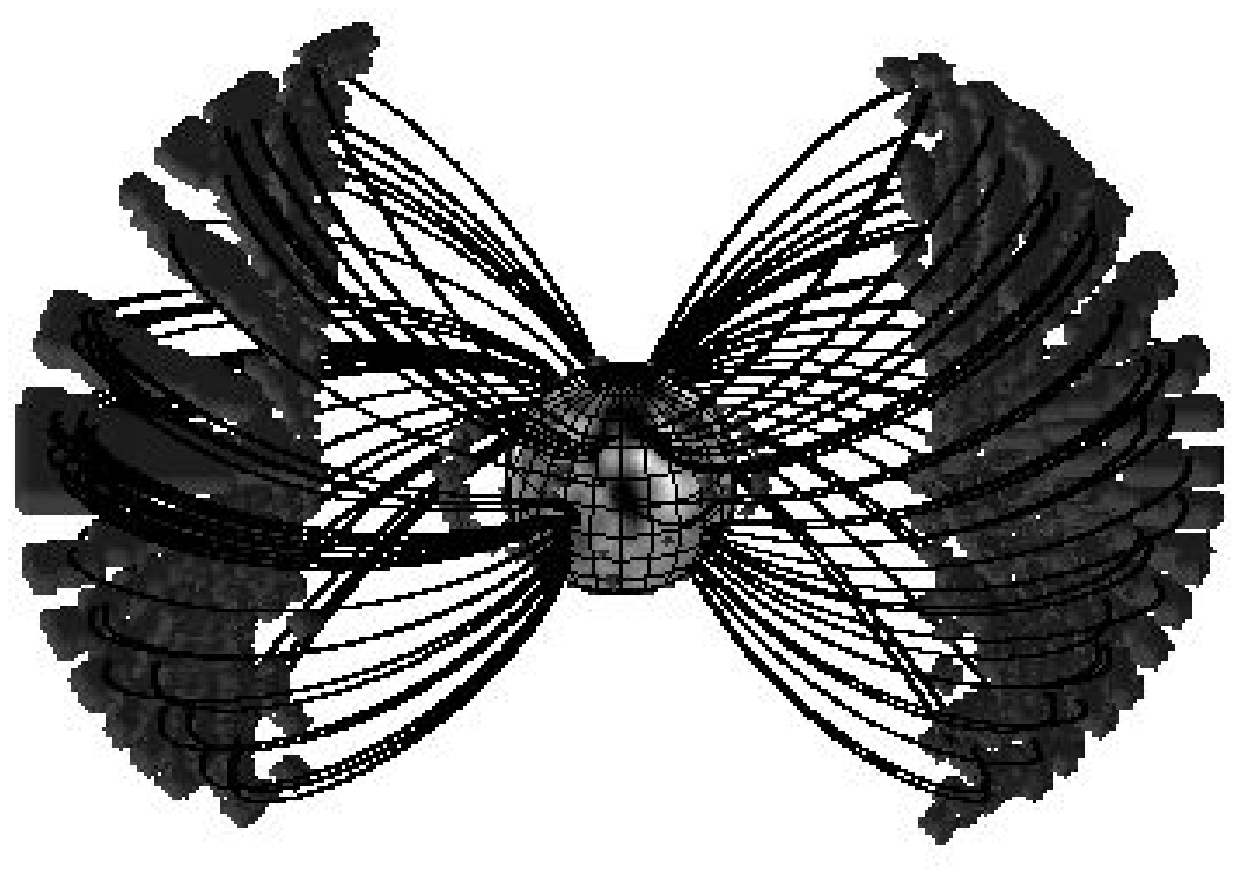}{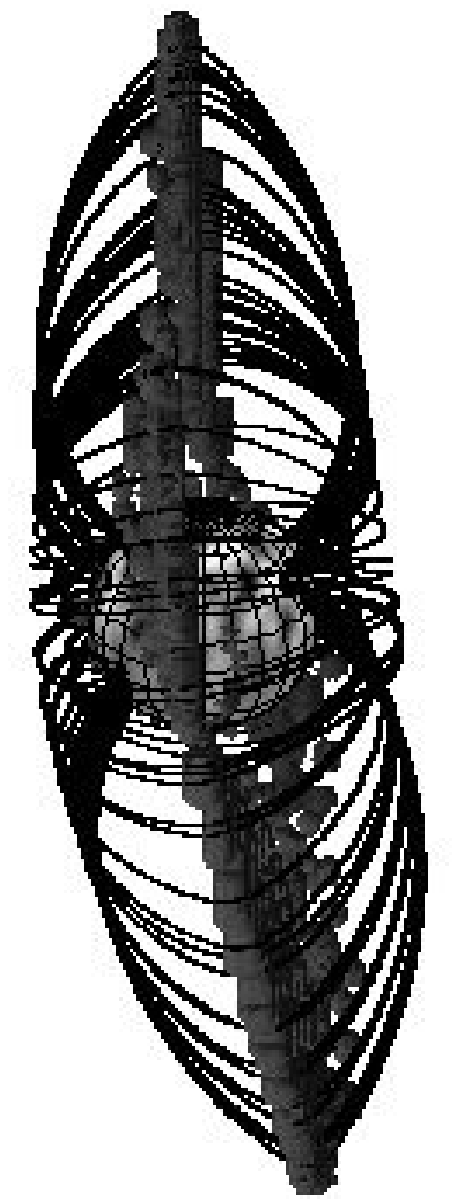}
\caption{Locations of mechanical equilibria showing where prominence
type complexes can be supported centrifugally. This model was 
computed using the coronal topology in Fig.5a 
(source surface, $R_s=5${\,\mbox{$\mbox{R}_*$}}). Phases 0.25 and 0.5 are pictured here.}
\label{fig:prom}
\end{figure*}

Our model of stable sheet-like surfaces are shown in Fig.~\ref{fig:prom}.
This plot shows that the main loci of equilibria where
prominences could form starts at about the Keplerian co-rotation radius
(2.6{\,\mbox{$\mbox{R}_*$}}) and extends out to the source surface ($R_s=5${\,\mbox{$\mbox{R}_*$}}). 
Note that although closed loops form over the pole, 
there are no stable equilibria here as there is no centrifugal support.
The prominences shown here would transit the stellar disk as 
the star rotates and could explain the observed H$\alpha$ absorption transients.
The model shown in Fig.~\ref{fig:prom} would suggest that prominences
can  form above 2.5{\,\mbox{$\mbox{R}_*$}}\ up to the source surface, 
but that they should be confined to two small ranges of longitude 
(approximately $180^{\circ}$-$200^{\circ}$ and $350^{\circ}$-$20^{\circ}$).
However, Donati et al. (1999) measure prominence-type complexes lying between 
2.5-5{\,\mbox{$\mbox{R}_*$}}\ at almost all observed longitudes 
using this same dataset. 
Therefore the model with $R_s=5${\,\mbox{$\mbox{R}_*$}} cannot explain the wide
range of longitudes at which prominences are found to occur. In contrast,
a model with lower source surface would produce many more polarity reversals.
Therefore, a possible solution of the longitude problem is that 
$R_s=1.7${\,\mbox{$\mbox{R}_*$}} and the prominences are located above
the source surface at longitudes where the radial magnetic field changes sign 
(similar to the coronal streamers observed on the Sun). In this case the 
prominences are located in a nearly radial field and cannot be in static 
equilibrium. Further modeling is required to determine whether the observed
H$\alpha$ prominences can be explained in this way. 

A word of caution when comparing this model with observations is 
necessary here. 
The analysis of prominences using H$\alpha$ data carried 
out by Donati et al. (1999) suggests that coronal features evolve
on a timescale of under 2 days 
and hence much more quickly than timescales of surface features 
(which are thought to have timescales on the order of 2 weeks or more). 
The coronal model that we have computed is based on surface maps
that were obtained by combining data taken over one whole week.
Observations would suggest that the coronal field or at least the component
of the field that supports prominences is subject to considerable variability
over the course of a week.
This model also does not take the field in the starspots into account. 
The missing field is of unknown strength and polarity and, particularly,
if the field in the high latitude spots is unipolar it will flatten the
prominence sheet into an equatorial current sheet.

\section{Discussion}

The magnetic field analysis of the stellar surface indicates that there 
is a strong non-potential field encircling the polar spot 
on AB Dor in 1996 December. 
This azimuthal field in located in the penumbra of the polar spot and its 
detection may be affected by the reduced visibility of the magnetic
fields in the spotted regions. However, the observed azimuthal field
appears to be real. 

Schrijver \& Title (2001) model the effect that the 
increased magnetic flux injected onto the surfaces of solar-type stars 
would have on subsequent flux patterns. They find that  concentric
bands of radial field with opposite polarity form near the pole
of the star. This is a consequence of meridional transport of flux
and the increased amount of flux dissipating on longer timescales. 
While our maps do not show this pattern, if there is strong flux of
opposite polarities in the polar spot, meridional fields
would connect the alternating bands of radial field.
As the star rotates differentially, this horizontal field would be sheared
in the azimuthal direction. Therefore, this model can in principle
explain the origin of the azimuthal field.

\begin{figure}
\epsscale{1}
\plotone{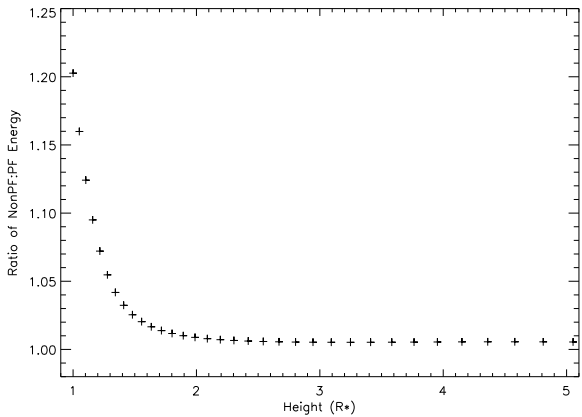}
\caption{The ratio of non-potential to potential magnetic field energy as a function of height. As this figure shows, the amount of free energy is greatest at the surface and drops quickly, levelling out  to under 1\% about 1{\,\mbox{$\mbox{R}_*$}}\ above the photosphere.}
\label{fig:freeen}
\end{figure}

Alternatively, there is preliminary evidence associating this azimuthal
feature with the epoch of observation. First results suggest that
the 1995 December images show no strong non-potential component and that
1998 January images have an even stronger non-potential component
than required by the 1996 December dataset. 
Could this non-potential component be related to the stellar activity cycle?
Does it switch direction over the course of the cycle? These are
questions that analysis of subsequent datasets will answer.
If this non-potential component is indeed real it is possible to
evaluate the amount of 
magnetic energy available to power flares and coronal mass ejections.
The amount of free energy integrated over the entire coronal volume is 14\%
of the potential-field energy. 
Fig.~\ref{fig:freeen} shows how the ratio of  
non-potential and potential magnetic energy densities depends on height. 
The amount of extra energy drops off with height from 20\% at the 
surface to well under 1\% at the source surface height (5{\,\mbox{$\mbox{R}_*$}}). 

Analysis of subsequent datasets will reveal how typical the 
features recovered in this paper are and if they evolve from year
to year. The method presented here promises to be an important tool
in probing the dynamo activity of rapidly rotating cool stars.

When modeling the coronal structure of the star it is necessary to 
define the point beyond which the  field is open and radial. 
In the model presented in Fig.~\ref{fig:prom}, this value has been set to 
5{\,\mbox{$\mbox{R}_*$}}\ as suggested by the presence of prominences at these 
distances. This model would apply if these prominences fit the model of 
slingshot-type prominences. 
Observations suggest that prominences form out to 5{\,\mbox{$\mbox{R}_*$}}, are stable 
over one rotation cycle but tend to have been ejected over a period of two days.
This instability may reflect the inherent variability in the coronal field 
of AB Dor. Surface features recovered on AB Dor are, by contrast, 
stable on timescales of over a week. 

If the source surface is at 1.7{\,\mbox{$\mbox{R}_*$}}\ as suggested by an analysis
of the gas and magnetic pressures in  AB Dor's corona (see Paper II),
the prominences are located beyond the source surface and their
equilibrium and stability cannot be described in terms of potential field models.
The magnetic field beyond the source surface
is nearly radial with current sheets 
separating regions with opposite sign of $B_r$.
According to this new model the prominences would be associated with 
these current sheets, but unlike in Fig.~\ref{fig:prom} they would
not be in a state if magnetostatic equilibrium.
Instead, the prominence plasma would slowly flow outward along the current
sheets with velocity significantly less than the rotational velocity.
This new model may explain the observation that prominences occur over a  broad 
range of longitudes.

\acknowledgments
The authors would like to thank the referee for comments that have
improved the final version of the paper. GAJH was supported by a
Harvard-Smithsonian Center for Astrophysics Postdoctoral Fellowship. ACC
acknowledges the support of a Senior Research Fellowship from the UK
Particle Physics and Astronomy Research Council. The UK Particle Physics
and Astronomy Research Council also funded research visits for GAJH that
aided the completion of this paper.

\end{document}